\documentclass[aps,prb,twocolumn,superscriptaddress,showpacs]{revtex4-1}
\usepackage{graphicx} 
\usepackage{epsfig}
\usepackage{amsmath}
\def\gapx{\lower 2pt \hbox{$\buildrel>\over{\scriptstyle{\sim}}$\ }}
\def\lapx{\lower 2pt \hbox{$\buildrel<\over{\scriptstyle{\sim}}$\ }}
\usepackage{amsmath}
\def\he4{$^4$He}
\def\he3{$^3$He}
\def\beq{\begin{equation}}
\def\eeq{\end{equation}}
\def\Am2{\AA$^{-2}$}
\begin{document}
\title{Low-density phases of $^3$He monolayers adsorbed on graphite}

\author{Michele Ruggeri}
\affiliation{SISSA Scuola Internazionale Superiore Studi Avanzati, 
Via Bonomea 265, I-34136 Trieste, Italy}
\author{Ettore Vitali}
\affiliation{Department of Physics, The College of William and Mary, 
Williamsburg, Virginia 23187}
\author{Davide Emilio Galli}
\affiliation{Dipartimento di Fisica, Universit\`a degli Studi di Milano, 
via Celoria 16, 20133 Milano, Italy}
\author{Massimo Boninsegni}
\affiliation{Department of Physics, University of Alberta, 
Edmonton, Alberta, Canada, T6G 2E1}
\author{Saverio Moroni}
\affiliation{CNR-IOM DEMOCRITOS Simulation Center, 
Via Bonomea 265, I-34136 Trieste, Italy}
\affiliation{SISSA Scuola Internazionale Superiore Studi Avanzati, 
Via Bonomea 265, I-34136 Trieste, Italy}

\date{\today}

\begin{abstract}
Quantum Monte Carlo simulations at zero temperature of a $^3$He monolayer 
adsorbed on graphite, either clean or preplated with $^4$He, unexpectedly 
point to a gas-liquid phase transition at a very low areal density of the 
order of 0.01\AA$^{-2}$.
This result stems from an essentially unbiased calculation of the ground
state energy for an infinite, defect-free substrate which interacts with
He atoms via  
a realistic potential, whereas 
the interaction between two He atoms includes two- and three-body terms. The sensitivity of the gas-liquid coexistence region on the model 
Hamiltonian employed is discussed.
\end{abstract}

\maketitle

\section{Introduction}
Thin films of either isotope of He adsorbed on substrates have been
widely studied experimentally as close realizations of ideal, strictly 
two-dimensional (2D) quantum systems with a coupling strength tunable by varying
the coverage. For example, specific heat and magnetization measurements
on the first and second adlayers of $^3$He on graphite
give very similar results,\cite{greywall,lusher,morhard} structural differences between successive
monolayers notwithstanding. Thus, these layers are deemed to provide a close realization of the 
2D model, allowing one to obtain experimentally estimates  of such fundamental properties of 2D 
Fermi liquid as the effective mass and  spin susceptibility. Likewise, the superfluid transition
of a $^4$He liquid adlayer has been predicted\cite{mct} and observed\cite{bishop,agnolet, tulimieri,luhman} to
adhere rather closely to the 2D (Kosterlitz-Thouless) paradigm.\cite{kt}

\begin{figure}[h]
\includegraphics[scale=0.60]{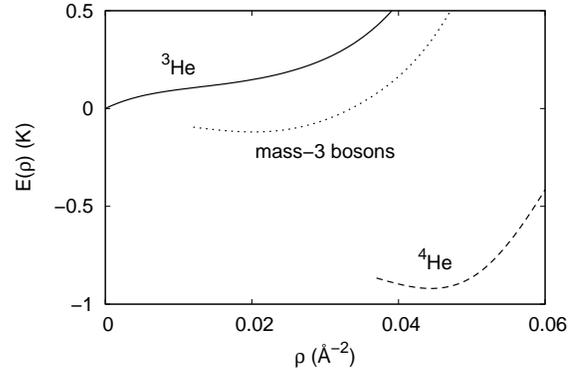}
\caption{
Ground state energy per atom $E(\rho)$ of 2D $^3$He (solid line),
$^4$He (dashed line), and fictitious mass-3 bosonic Helium (dotted line). 
The global minima
for $^4$He and mass-3 bosons locate the equilibrium densities of the 
self-bound liquid phases, respectively 0.044~\AA$^{-2}$ and 0.020~\AA$^{-2}$. 
For $^3$He there is neither self-bound liquid nor gas-liquid phase transition.
}
\label{fig_ideal2D}
\end{figure}

Fig. \ref{fig_ideal2D} shows  the zero temperature 
equations of state (energy per particle $E$ 
versus density $\rho$) of 2D $^4$He, $^3$He and a fictitious bosonic version of spin-zero $^3$He, computed in this work as
decribed below. The results are in agreement with those of previous, comparable calculations,\cite{whitlock,campbell,miller,um,grau,nava} with only minimal
differences arising from the use of slightly different interatomic potentials. As one can see,
$^4$He features a self-bound liquid 
phase of equilibrium density $\rho_0 = 0.044$ \AA$^{-2}$,  below which the system forms puddles; 
on the other hand, $^3$He remains in a homogeneous 
fluid phase from the crystallization density all the way down to the ideal 
gas limit. 
The difference between $^3$He and $^4$He arises because of the higher kinetic energy of $^3$He, which is due 
in part to its lighter mass, and in part to Fermi statistics -- the latter being at the root of the qualitatively different shape of the
$E(\rho)$ curve for $^3$He, compared to that for $^4$He and bosonic spin-zero $^3$He, also a self-bound liquid.

Although no liquid phase of $^3$He exists in purely 2D, there are reasons to expect
that it may in quasi-2D, i.e., in a thin film adsorbed on a suitable substrate. This is because 
adatoms experience two distinct effects which  favor the formation of a liquid phase of $^3$He
(and enhance the stability of that of $^4$He), compared to the strictly 2D case.
The first is quantum delocalization in the direction perpendicular to the substrate,
which acts to soften the hard-core repulsion of the helium interatomic 
potential at short distances; the second is  substrate corrugation, which
effectively increases the mass of the Helium atoms.\\ \indent
For example, on weakly attractive substrates such as those of alkali metals, 
where delocalization is significant and whose corrugation can be regarded as negligible, on account of the 
relatively large equilibrium distance of the atoms from the surface, 
$^3$He has been predicted\cite{quasi2d} to undergo a
phase transition between a gas and a thermodynamically stable 
(albeit not self-bound) liquid, signalled by a local minimum in 
the equation of state $E(\rho)$. \\ 
The stability of the liquid phase
weakens for increasingly attractive  
substrates, suggesting that $^3$He may not condense into a stable liquid on 
a substrate as strong as graphite. On the other hand, as adatoms move closer to the
surface, the enhancement of their effective mass arising from substrate corrugation may again
underlie a thermodynamically stable liquid phase. More generally,  the existence of a gas-liquid transition of a quasi-2D 
adsorbed $^3$He film hinges on the  interplay of a number of 
subtle, substrate-dependent effects.
\\ \indent
Although the phase diagram of Helium adsorbed on graphite has been extensively investigated experimentally, 
there is no general consensus about the existence of one or more fluid phases, even in the  first  $^3$He adlayer. 
The prevailing opinion has been that  $^3$He should form a 2D gas at low coverage,  directly crystallizing into a registered solid as
coverage is increased; however, there exist  conflicting theoretical predictions\cite{ab,miller,campbell,bjl}
and  the lack of data at sufficiently low temperatures and coverage 
to provide a definitive experimental answer.\cite{godfrin} 
The issue has received 
renewed consideration after a recent measurement of the specific 
heat, whose linear dependence on the density at very low coverages has been
interpreted in terms of puddles of self-bound liquid. Interestingly,
similar results were reported for the first, second and third 
adlayers of $^3$He on graphite.\cite{sato}

In this work, we use quantum Monte Carlo simulations to
calculate with high accuracy the $T=0$ low-density equation of 
state (EOS) of $^3$He on graphite (G), and on graphite preplated 
with a solid monolayer of $^4$He (G4He). In both cases the
compressibility becomes negative in a small density region, 
implying a pressure-induced phase transition between a 
finite-density gas and a liquid.
\\ \indent
This result falls short of  providing a direct, unambiguous quantitative confirmation of the findings of Ref. \onlinecite{sato}, which 
is to be expected given the limitations of the (still relatively simplified) theoretical model utilized here. Nevertheless, it clearly lends support to the idea that a stable liquid phase may exist, one which in the case of G is crucially underlain by the substrate corrugation.
The reminder of this article is organized as follows: in Sec. \ref{sec_methods} we describe the mathematical model and the computational methodology utilized in this work; Sections \ref {sec_results} and \ref{sec_details} are devoted to a thorough illustration of our results, and in particular in Sec. \ref{sec_details} we offer specific, quantitative details, for the purpose of facilitating the task of others who may 
wish to reproduce our results. We outline our conclusions and prospects for further studies in Sec. \ref{discussion}.

\section{Models and method}
\label{sec_methods}
We consider a system of $N$ $^3$He atoms in the presence of a planar substrate,
enclosed in a cell of sides $L_x, L_y$ and $L_z$ with periodic boundary 
conditions. The quantum-mechanical many-body Hamiltonian is the following:
\begin{equation}\label{hamiltonian}
H=-\lambda\sum_{i=1}^N\nabla_i^2+V(R)+\sum_{i=1}^N U({\bf r}_i),
\end{equation}
where $\lambda=8.0417$ K\AA$^{-2}$,  $R=\{{\bf r}_1,\ldots,{\bf r}_N\}$ are the
coordinates of the $^3$He atoms, $V(R)$ is the He-He interaction
and $U({\bf r})$ is the He-substrate interaction. The areal density of
the system is defined as $\rho=N/(L_xL_y)$, with the $z$ axis perpendicular
to the substrate.
As we aim to model the system as realistically as possible, in $V(R)$ 
we include the highly accurate, first-principle SAPT2 pair 
potential\cite{korona} and the Axilrod-Teller-Muto three-body 
potential,\cite{at,nota3} which at low density is by far the most 
important term beyond pair-wise interactions.\cite{pederiva}
Concerning the He-substrate interaction U({\bf r}), 
for $^3$He directly adsorbed on G we use the anisotropic potential 
of Carlos and Cole\cite{cc} (CC) which accounts for substrate corrugation and has been adopted in most computational
studies of $^4$He adsorbed on graphite.\cite{pierce,pierce2,pierce3,corboz}
For $^3$He adsorbed on G4He instead, we derive an effective potential 
which is significanltly weaker than for G and only depends on the $z$ 
component of ${\bf r}$. Since the procedure is somewhat elaborate,
we defer its discussion to Sec.~\ref{sec_details} to avoid 
diverting from the main results of this work.
\\ \indent
The ground state energy of the model Hamiltonian (\ref{hamiltonian})
is calculated by means of fixed-node Diffusion Monte Carlo
(FNDMC), \cite{reynolds82,umrigar93,mitas_rmp} a 
widely used quantum Monte Carlo technique which gives very accurate
energy upper bounds by projecting the lowest-energy fermion
eigenstate of the Hamiltonian with the same nodes as a trial wave
function $\Psi$. Specifically, FNDMC simulates the imaginary-time 
evolution of the system in configuration space through a branching 
random walk of a large (ideally infinite) number of walkers, subjected
to the constraint that the nodes of $\Psi$ never be crossed. 
Beside (i) the finite size of the system, FNDMC is subject to errors from 
(ii) the time discretization of the random walk, (iii) the finite 
number of walkers, and (iv) the fixed-node constraint.
The errors due to (i--iii) can be estimated varying the number 
of particles, the time step, and the number of walkers, and eliminated 
by extrapolation whenever deemed non-negligible. The only uncontrolled error, namely
the fixed-node approximation, depends on the nodal structure of $\Psi$. 
\\ \indent
The trial many-body wave functions used in this work have the Jastrow-Slater form
$\Psi=\exp(-U)D_\uparrow D_\downarrow$. 
The Jastrow factor $U=U_1+U_2+U_3$ contains standard two- and three-body 
correlations between Helium atoms,\cite{schmidt} as well as one-body 
Helium-substrate correlations:
\begin{eqnarray}
\label{eq_psi}
U_1&=&\sum_i f(z_i)+\sum_{ij}g(|{\bf r}_i-{\bf t}_j|),\nonumber\\
U_2&=&\sum_{i<j}u(r_{ij}),\nonumber\\
U_3&=&\sum_{j\neq i}\xi(r_{ij})({\bf r}_i-{\bf r}_j)\cdot
      \sum_{l\neq i}\xi(r_{il})({\bf r}_i-{\bf r}_l).
\end{eqnarray}
The second term in $U_1$ is a radial correlation between each Helium atom
and an array of sites placed at positions ${\bf t}_j$, shifted by an amount
$d$ above each Carbon atom of the graphite surface. It is used only for the
corrugated substrate.
\\ \indent
In the Slater determinant for
$\alpha$-spin atoms, namely $D_\alpha={\rm det} (e^{-i{\bf k}_l\cdot{\bf q}_j})$, 
the one-particle orbitals are plane waves with the $N/2$ smallest wave 
vectors of the form $(2\pi n/L_x,2\pi m/L_y)$
and the 2D coordinates
\begin{equation}
{\bf q}_j={\bf s}_j+\sum_{l\neq j}\eta(s_{jl})({\bf s}_j-{\bf s}_l),
\end{equation}
with ${\bf s}_j=(x_j,y_j)$, include backflow correlations\cite{schmidt}
in the nodal structure of $\Psi$. Based on a comparison
between fixed-node and nominally exact Transient Estimate (TE) 
results\cite{nava} for strictly 2D $^3$He, backflow wavefunctions are 
expected to be very accurate, at least at low density.
\\ \indent
All the radial functions $u$, $\xi$, $\eta$,
$f$ and $g$ are parametrized as linear combinations of powers of their
argument. The coefficients, a couple of negative exponents,
and the shift $d$, for a total of about 30 variational parameters,
are optimized using correlated sampling.\cite{fantoni}
The FNDMC energies utilized in the results presented in the Sec. \ref{sec_results}
include corrections for all sizeable sources of bias, including an estimate 
of the fixed-node error,\cite{nava,superbf} as explained in detail in 
Section~\ref{sec_details}.

\section{results}
\label{sec_results}
\begin{figure}[h]
\includegraphics[scale=0.60]{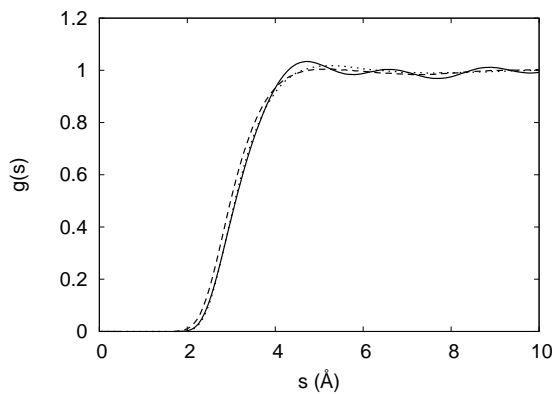}
\caption{
Pair correlation function $g(s)$ 
at $\rho=0.020$~\AA$^{-2}$
for $^3$He adsorbed on G (solid line) and G4He (dashed line), 
and for strictly 2D $^3$He (dotted line). The statistical error is smaller
than the thickness of the lines.
}
\label{fig_gofr}
\end{figure}

\begin{figure}[h]
\includegraphics[scale=0.60]{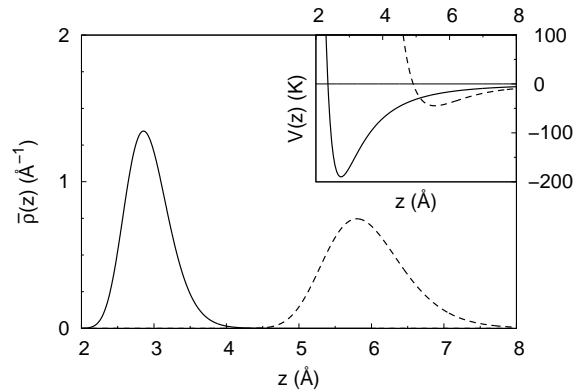}
\caption{
Normalized density profile $\bar\rho(z)$ 
for $^3$He adsorbed on G (solid line) and G4He (dashed line). 
The data refer to an areal density $\rho=0.020$~\AA$^{-2}$, 
but $\bar\rho(z)$ is nearly independent of $\rho$.
The statistical error is smaller
than the thickness of the lines.
The inset shows the $^3$He--substrate potential for G (lateral average
of the anisotropic potential, solid line) and G4He (dashed line).
}
\label{fig_rhoz}
\end{figure}

\begin{figure}[h]
\includegraphics[scale=0.60]{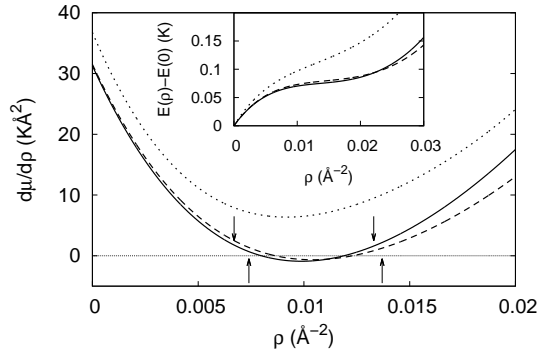}
\caption{
Derivative of the chemical potential with respect to the density
for $^3$He adsorbed on G (solid line) and G4He (dashed line), 
and for strictly 2D $^3$He (dotted line).
Negative values signal instability to phase separation into a gas and
a liquid phase For G and G4He, but not for 2D. 
Downward (upward) arrows indicate the boundaries of the 
coexistence region for G (G4He).
The inset shows the EOS (energy per particle vs. density).
}
\label{fig_mup}
\end{figure}

The structural properties of $^3$He adsorbed either on
G or on G4He suggest a close similarity between the adsorbate
and the strictly 2D system.
The vertically integrated pair correlation function $g(s)$, where $s=\sqrt{x^2+y^2}$ , 
is shown in Fig.~\ref{fig_gofr}. For G, apart from the oscillations
induced by the surface corrugation, $g(s)$ is hardly distinguishable
from the 2D case. In particular, the steep rise starting at $s\sim 2$~\AA,
which defines the excluded area around each particle 
induced by the short-range He-He repulsion, is
nearly identical for G and 2D. 
This is consistent with the quasi-2D character of the film.
The normalized density profile 
$\bar\rho(z)=\rho(z)/\int \ dz\ \rho(z)$, with $\rho(z)=\int \ dx\ dy\ \rho(x,y,z)$, is shown in Fig.~\ref{fig_rhoz}.
On a substrate, two Helium atoms can approach the same $(x,y)$ coordinates
if their distance along $z$ is larger than the excluded volume radius of
about 2~\AA. This effectively results in a
softening of the short-range repulsion with respect to the bare
interparticle potential. However, as shown in Fig. \ref{fig_rhoz}, 
for G the full extent of the density profile barely reaches 
2~\AA, thus leaving little room
for delocalization to soften significantly the short-range repulsive part of the two-body He-He  potential, in turn
reducing the size of the correlation hole in $g(s)$.
\\ \indent
For G4He, apart from the obvious shift of the peak position to the second 
adlayer (the first one being occupied by the $^4$He solid), the
density profile is almost twice as broad as for G, 
on account of the shallower He-substrate
potential shown in the inset of Fig.~\ref{fig_rhoz}.  
Nonetheless, G4He is still a relatively strong substrate, in fact slightly stronger than 
Mg,\cite{quasi2d} resulting in a narrower $\bar{\rho}(z)$,  a wider correlation hole
 in $g(s)$, and a deeper He-substrate potential, albeit not by a large amount.
\\ \indent
In Ref.~\onlinecite{quasi2d}, a series of substrates were studied, the strongest being
Mg. A local minimum was found in the EOS of $^3$He
adsorbed on weak alkali metal substrates, but not on Mg,
so that none is expected on the even stronger G4He or G.
This is indeed the case, as shown in the inset of Fig.~\ref{fig_mup}. 
However, in a computer simulation of a system of finite size,  the occurrence of a gas-liquid phase transition  in a given range 
range of density is also signalled by a {\it negative} value of $d\mu/d\rho$, where $\mu(\rho)=E(\rho)+\rho\ (dE(\rho)/d\rho)$ is the chemical
potential. This is a weaker condition than the presence of a local minimum of $E(\rho)$.
\\ \indent
 From a polynomial fit to the FNDMC energies
calculated at several densities with statistical errors of 1.5~mK or less, we 
obtain the quantity $d\mu/d\rho$, shown in Fig.~\ref{fig_mup}. It is clearly seen to change
sign for both G and G4He in a small density interval around 0.01~\AA.
Having determined that a phase transition occurs 
on both G and G4He substrates (for the models considered),
we locate the coexistence region by searching 
each EOS for two points on either side of the interval of negative 
$d\mu/d\rho$ with the same pressure and chemical potential.
The results are listed in Table~\ref{table_coe}.
\begin{table}[h]
 \begin{tabular}{|l|c|c|c|c|} \hline
  & $\rho_{\rm min}$ (\AA$^{-2}$) & $\mu'_{\rm min}$ (K\AA$^2$) & $\rho_{\rm L}$ (\AA$^{-2}$) & $\rho_{\rm V}$ (\AA$^{-2}$) \\ 
  \hline
G    &  0.0099(3) & -0.89(25) & 0.0067(6) & 0.0133(7) \\
G4He &  0.0104(1) & -0.65(11) & 0.0074(2) & 0.0137(4) \\
    \hline 
 \end{tabular}
\caption{
Location $\rho_{\rm min}$ and value $\mu'_{\rm min}$ of the minimum
of the derivative of the chemical potential with respect to the density,
liquefaction density $\rho_{\rm L}$ and vaporization density $\rho_{\rm V}$
of $^3$He adsorbed on G and G4He. Statistical errors on the last
digit(s) are given in parenthesis.
}
\label{table_coe}
\end{table}
Note that the coexistence regions are wider than 
the intervals of negative compressibility.
In the intervening left and right density windows
the gas and liquid phases, respectively, are {\it metastable}.
\\ \indent
While the minimum of $d\mu/d\rho$ for the 2D model has a clear
bearing on the location of the coexistence regions for both G 
and G4He, the unexpectedly close similarity of the $^3$He EOS 
on the two substrates  should be regarded as largely fortuitous. Indeed, delocalization 
effects are different for G and G4He, but the corrugation, 
which is significant for G and negligible for G4He (see 
Section~\ref{sec_details}), happens to  compensate nearly perfectly for that
difference.
\\ \indent
We conclude this Section by mentioning,
in relation to the tentative phase diagrams discussed in Ref.~\onlinecite{bjl}, 
that the ground state energy per $^3$He atom
in the ${\sqrt 3}\times{\sqrt 3}$ solid phase commensurate with G 
is found in this calculation to be 
$E(\rho)-E(0)=0.62(1)$~K, much higher than 
in the gas or liquid phases at coexistence.

\section{details of the calculation}
\label{sec_details}
In this Section we discuss the sources of bias mentioned in Section
\ref{sec_methods} and the calculation of the interaction potential
between a $^3$He atom and the G4He substrate employed in Section
\ref{sec_results}.
\subsection{Sources of bias}

\begin{figure}[h]
\includegraphics[scale=0.60]{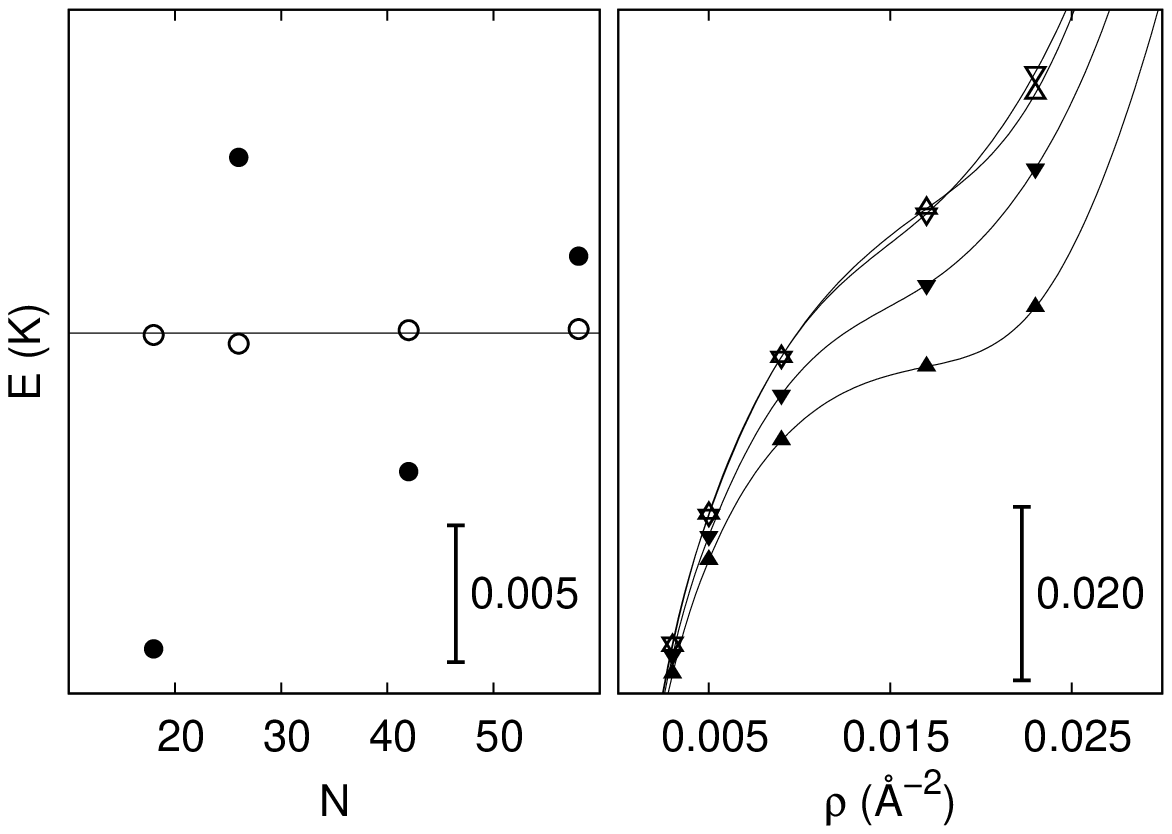}
\caption{
{\it Left.} FNDMC energy per particle for 2D $^3$He at $\rho=0.09$~\AA$^{-2}$
as a function of the number of particles: raw data (filled symbols)
and values including the finite size correction of Eq.~(\ref{eq_size})
(empty symbols). The line is a constant fit to the size-corrected
data.\\
{\it Right.} FNDMC energy per particle for $^3$He on G4He calculated with $N=18$
(upward triangles) or $N=42$ (downward triangles) as a function of density: 
raw data (filled symbols) and values including the finite size correction 
taken from the strictly 2D system (empty symbols).
The lines are spline interpolations.
In both panels the statistical errors are smaller than the symbol size.
}
\label{fig_size}
\end{figure}

{\em Finite size effects}. All the results presented in Section 
\ref{sec_results} are based on FNDMC simulations of systems of $N=18$
$^3$He atoms\cite{note18}. For the 2D system the finite size error
has been estimated from simulations of larger systems.
An example is shown in Fig.~\ref{fig_size} (left) for $\rho=0.09$~\AA$^{-2}$.
The filled symbols indicate the FNDMC energies obtained as a function of
$N$, with a tail correction calculated assuming $g(r)$=1 for $r$ larger
than half the side of the simulation cell. The scatter of the data
is strongly reduced by applying the correction\cite{tanatar}
\begin{equation}
E(\infty)-E(N)=\alpha[T_0(\infty)-T_0(N)],
\label{eq_size}
\end{equation}
where $T_0(N)$ is the kinetic energy of a non-interacting system
of $N$ particles and $\alpha$ is a fit parameter:\cite{notafit} this correction
brings the data onto the empty symbols, which lie on a smooth curve
(in fact a constant to good accuracy) which is our estimate 
of the energy in the thermodynamic limit. Figure~\ref{fig_size} (right)
shows that the 2D size correction applied to the adsorbate on 
G4He (the less close to the 2D limit) gives essentially indistinguishable
results for $N=18$ and $N=42$. We thus conclude that the size correction
of Eq.~(\ref{eq_size}) with the value of $\alpha$ determined for the 2D
case can be safely used for the adsorbates as well.
No significant finite size effects were detected on the structural properties
at the densities studied in the present work.

\begin{figure}[h]
\includegraphics[scale=0.60]{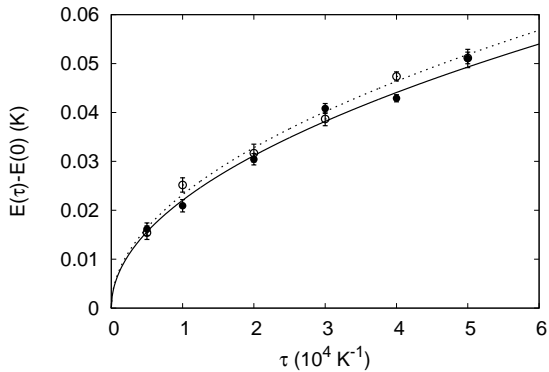}
\caption{
Dependence of the energy $E$ on the time step $\tau$ for $^3$He on G.
The filled (empty) circles with error bars are FNDMC data, and the solid
(dotted) line is a fit in $\tau^{1/2}$, for $\rho=0.005$ (0.020) \AA$^{-2}$.
}
\label{fig_tau}
\end{figure}

{\em Time step error}. For both substrates we performed simulations
with several values of the time step $\tau$ spanning an order of magnitude.
We determined that the EOS $E(\rho)-E(0)$ is not affected by the time
step error, within an accuracy of $\lesssim 1$~mK, using 
$\tau=5\times 10^{-4}$~K$^{-1}$
for G4He and $\tau=10^{-4}$~K$^{-1}$ for G. In particular, this rather 
conservative choice for G is due to the apparent $\tau^{1/2}$ dependence of the
energy\cite{notatstep}, shown in Fig.~\ref{fig_tau}. For G4He and 2D a much weaker, linear
time step dependence is observed.

{\em Population control bias}. This has been shown to be a potentially serious, often overlooked problem in 
DMC simulations.\cite{hinde,nemec,bm}
For the low densities and small systems
considered here, however, the error due to the finite number of walkers is not a concern.
We used 6400 walkers, after verifying that this is far more than enough
to eliminate this source of bias within the accuracy of the present
calculations.
\\ \indent
{\em Fixed-Node approximation}. The FN error can be estimated
from published TE results for the 2D case. From Ref.~\onlinecite{nava}
a quadratic dependence on the density can be inferred, and 
Ref.~\onlinecite{superbf} reports a FN error of $\sim$0.046~K at 
${\bar\rho}=0.060$~\AA$^{-2}$ using a trial function 
of quality comparable to those used here. We thus add a correction
$-0.046(\rho/{\bar\rho})^2$~K to our FNDMC energies, for both the 2D
and the adsorbate systems. 
\\ \indent
{\em Mixed estimators}. For quantities other than the total energy,
the ``mixed'' estimators directly obtained\cite{mitas_rmp} in DMC have a bias
linear in the error of the trial function. The structural properties
presented in this work are ``extrapolated'' estimators,\cite{mitas_rmp}
whose bias is quadratic in the error of the trial function.
A small difference between extrapolated and mixed estimators is often
considered as a qualitative indication of small residual bias.
With the trial functions used in this work,
this difference is very small for the pair distribution function
and barely discernible on the density profile.
\subsection{The He-G4He potential}
\begin{figure}[h]
\includegraphics[scale=0.70]{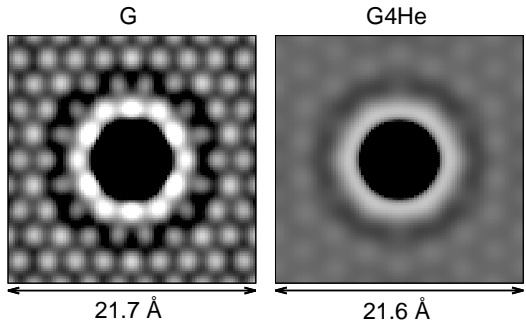}
\caption{
Pair distribution function $g(x,y)$ for $^3$He
adsorbed on G at $\rho=0.038$~\AA$^{-2}$
and on G4He at $\rho=0.038$~\AA$^{-2}$.
The grey-scale range, restricted to 0.94--1.19,
emphasizes the long-range oscillations 
(the highest peaks for G reach 1.28).
}
\label{fig_gofxy}
\end{figure}

In a series of preliminary calculations we modeled the G4He substrate
with ``active" (i.e., explicitly simulated) $^4$He atoms at a density $\rho_{\rm 4He}=0.114$~\AA$^{-2}$,
close to the coverage where promotion to the second layer begins.\cite{lusher,corboz}
In the Jastrow factor of the trial function, namely Eq.~(\ref{eq_psi}),
a Nosanow term \cite{nosanow} was added to tie the $(x,y)$ coordinates of 
the $^4$He atoms to the sites of a triangular lattice, and 
different factors $f_3(z)$ and $f_4(z)$ were used in Eq.~(\ref{eq_psi}) to 
describe localization in $z$ for $^3$He and $^4$He, as well as different 
pair pseudopotentials $u_{33}$, $u_{34}$ and $u_{44}$.
The corrugation of the graphite surface was not expected to have
any significant effect on the $^3$He atoms confined to the second
monolayer by the intervening incommensurate $^4$He crystal. Therefore
the He-graphite interaction was described by the laterally averaged CC
potential.
\\ \indent
In the DMC simulation of a multicomponent system, the energy of a 
single component, such as the $^3$He EOS of specific interest here, 
is obtained as a biased mixed estimator.  Thus, its projected value is influenced by all the terms in the trial function
-- unlike the total energy  which only depends on the nodal structure.
As a result, statistical fluctuations in the optimal variational
parameters induce enough scattering in the density dependence
of the $^3$He energy to prevent us from extracting
the compressibility by numerical differentiation of the EOS. 
Thus, even neglecting the issue of the bias in the
mixed (or extrapolated) estimator of the $^3$He energy,
the use of this ``full'' model turns out to be impractical.
This prompts for a replacement of the $^4$He layer by a 
rigid effective potential. To this purpose, we retain
useful information on structural properties from the full
model, such as the $^3$He density profile of Fig.~\ref{fig_rhoz} 
and the pair distribution function of Fig.~\ref{fig_gofxy}. In particular
the density profile shows that $^3$He floats atop the $^4$He
solid, with minimal excursions of atoms of either species in the
monolayer occupied by the other, suggesting that a rigid potential
is a sensible choice. 

\begin{figure}[h]
\includegraphics[scale=0.60]{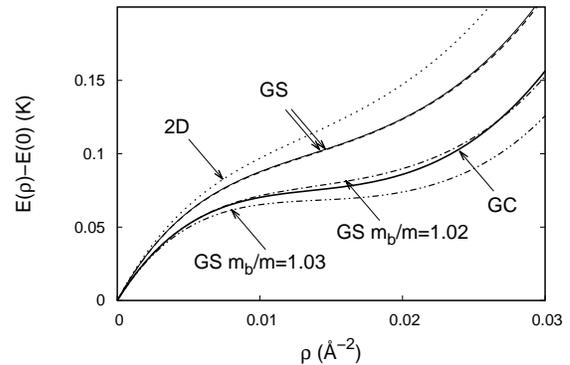}
\caption{
EOS for $^3$He adsorbed on G
with various He-substrate potentials and/or effective band masses. 
GC (thick solid line) refers to the corrugated G substrate discussed in
Section \ref{sec_results}. GS denotes smooth-substrate models, with either
the potential of Eq.~(\ref{eq_schrodinger}) (dashed line), or
the laterally averaged version of the CC potential (thin solid line, 
dash-dotted line and dash-double-dotted line). The value $m_b$ of the effective 
band mass, whenever different from the bare mass $m$, is as indicated.
The EOS for 2D $^3$He (dotted line) is also reported for reference.
}
\label{fig_eos}
\end{figure}

Making the further assumption of a smooth G4He substrate, we solve 
for the potential $U(z)$ the one-dimensional Schr\"odinger equation of 
a single $^3$He atom with a known density profile,
\begin{equation}
U(z)=\frac{\hbar^2}{2m}\psi^{-1}(z)\frac{\partial^2\psi(z)}{\partial z}+C,
\label{eq_schrodinger}
\end{equation}
where $\psi(z)$ is the square root of $\rho(z)$ 
calculated with the full model and $C$ is a constant.
The result for G4He is shown in the inset of Fig.~\ref{fig_rhoz}.
We assess the accuracy of this procedure on the G substrate,
where the anisotropic CC potential and its smooth, laterally 
averaged version are both available.\cite{cc} The EOS calculated 
with the potential of Eq.~(\ref{eq_schrodinger}) (with the
density profile of a $^3$He atom on the corrugated G substrate) 
is nearly identical to that calculated with the smooth CC potential, 
as shown in Fig.~\ref{fig_eos} by the dashed and the thin solid lines,
respectively.
\\ \indent
Of course a smooth substrate model constructed in this way misses
the corrugation effects. For the EOS of $^3$He on G 
they are significant, as seen in Fig.~\ref{fig_eos} by comparing
the results corresponding to smooth (GS) and corrugated (GC) substrates.
They can be included to good accuracy into the smooth model by replacing
the bare mass $m$ of the adsorbate with an effective band
mass $m_b$: figure~\ref{fig_eos} shows that the best agreement with the
GC substrate is obtained using $m_b/m=1.02$, somewhat smaller than
the value $m_b/m=1.03$ given in Ref.~\onlinecite{cc}.
Incidentally, this analysis shows that the difference between the 
2D system and the adsorbate on G is due more to the corrugation of
the substrate than to the delocalization of the adsorbate along $z$.
In order to assign an effective band mass value for the G4He substrate
in the lack of a reliable EOS for the full model,
we compare the two-dimensional pair distribution functions $g(x,y)$ of
$^3$He on G and G4He in Fig.~\ref{fig_gofxy}, whose long-range behavior
is representative of the density modulation of the adsorbate induced
by the corrugation of the substrate. This modulation
is stronger on G than on G4He by about a factor ten,
and assuming a similar ratio between the two substrates for other 
corrugation effects as well, we expect that for G4He the value
of $m_b$ be larger than $m$ by only a few parts in a thousand,
inducing a difference between
the EOS of the smooth and the full model ten times smaller than the
difference between GS and GC in Fig.~\ref{fig_eos}.
We consider this effect negligible and we treat G4He as a smooth
substrate using the bare mass for the $^3$He atoms.

\section{discussion}\label{discussion}
\begin{figure}[h]
\includegraphics[scale=0.60]{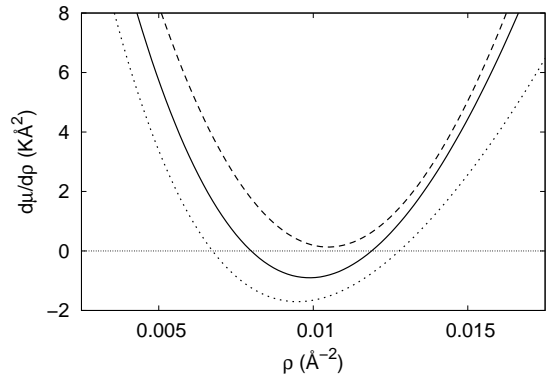}
\caption{
Derivative of the chemical potential with respect to the density
for $^3$He adsorbed on G, for three different model Hamiltonians.
The first (solid line) is the model of Section~\ref{sec_methods} 
(same as for the solid line in Fig.~\ref{fig_mup}).
The second (dashed line) has the He-He SAPT2 plus Axilrod-Teller 
potential replaced by the HFDHE2 Aziz potential. The third
(dotted line) has the CC anisotropic He-G potential replaced by
the smooth potential of Ref.~\onlinecite{bjl0} and the bare mass
of the He atoms replaced by an effective band mass $m_b=1.02 m$.
}
\label{fig_mup2}
\end{figure}

The key result of this work, namely the change of 
sign of $d\mu/d\rho$ shown in Fig.~\ref{fig_mup}, requires a sufficiently accurate calculation
in order to be established with reasonable confidence.  Although we believe that all sources of bias from
the methodology adopted here are well under control (see Section~\ref{sec_details}), 
we need to  stress that $d\mu/d\rho$ which results from differentiating twice a 
fit to FNDMC energies calculated with statistical errors of the order of 
a mK, goes below zero by a few standard deviations. 
We hardly feel comfortable with the assumption that the employed
model (see Section~\ref{sec_methods}), albeit of state-of-the-art
level, guarantees an accuracy on the energy of the order of 1 mK
throughout the relevant density range. 
\\ \indent
In order to substantiate this caution,  
we show in Fig.~\ref{fig_mup2} that different estimates of the 
model Hamiltonian lead to significant differences in the result.
This is the case even for the He-He interaction, arguably the most favorable case
for a force-field potential. Indeed, using the HFDHE2 Aziz potential\cite{aziz} 
instead of the SAPT2 potential supplemented with the Axilrod-Teller-Muto
three-body term, the density range of negative $d\mu/d\rho$, if any,
is within the statistical noise.\cite{nota} The HFDHE2 is a phenomenological pair 
potential which yields a fairly good equation of state\cite{kalos} for bulk $^4$He
in 3D. On the other hand, the SAPT2 potential comes from a very accurate quantum chemistry
calculation of the He dimer; in conjunction with the dominant three-body term, it is expected to afford greater 
accuracy than the HFDHE2, especially  at low density.\cite{pederiva} \\ \indent
It is important to note that the level of accuracy of either the 
HFDHE2 or the SAPT2 (or other modern He-He potentials) is {\it far} greater than that
of  {\it any} known  He-graphite potential.
In order to obtain a semiquantitative assessment of the amount of variation that the use of a different microscopic model
for the He-graphite interaction could entail,  we performed here a few calculations replacing the CC potential with 
the smooth He-graphite
interaction of Ref.~\onlinecite{bjl0}, with an effective $^3$He 
band mass $m_b=1.02 m$ (see Section \ref{sec_details}); to our relief,  the minimum value of $d\mu/d\rho$ changes
``only'' by $\sim$100 percent -- this time in the direction of  greater width of the region of negative $d\mu/d\rho$  (see Fig.~\ref{fig_mup2}).  It should be mentioned in fairness that  such a relatively  limited effect arising from the replacement 
of the He-graphite interaction
may be to some extent accidental; in general, it is clear that
a large uncertainty on our results stems from the He-graphite interaction.
Indeed, there are objective limitations to the accuracy that one can achieve in describing this system by means of static potentials. 
In particular, the interaction between two He atoms is itself subjected to screening 
effects from the graphite substrate.\cite{vidali} The estimated effect,
larger than the differences between different versions of the He-He pair potentials,\cite{korona,aziz} 
can have an influence on the delicate balance between competing phases\cite{bruch,vranjes}; 
in our case, it would tend to shrink or suppress the gas-liquid coexistence region 
(at least for the G substrate, where the $^3$He adatoms are closer to the Carbon atoms).
A further source of uncertainty is the motion
of the Carbon atoms\cite{gordillo}, not considered in our model Hamiltonian.
Nonetheless, it seems a safe conclusion that the phase diagram of
$^3$He on G or G4H4 is to say the least on the edge of  featuring a gas-liquid
coexistence region around $\rho=0.01$~\AA$^{-2}$.
\section{Conclusions}
We have performed state-of-the-art computer simulations, based on the most realistic model of the system of interest,
in order to address the question of the existence of a gas-liquid phase transition of a $^3$He monolayer adsorbed on 
a graphite substrate, either bare or preplated with $^4$He.  Our results, while not providing direct theoretical validation to the contended
experimental observation of such a phase, reported in Ref. \onlinecite {sato}, nonetheless generally support its existence, in a region of coverage close to the experimentally relevant one. \\ \indent
We have attempted to offer as extensive as possible a discussion of the limitations of this calculation; as mentioned above, the most serious source of uncertainty is the microscopic model utilized. 
\\ \indent
We note that the generally unexpected phase transition characterized 
in Table~\ref{table_coe} describes the liquefaction of a finite-pressure gas
on an infinite, defect-free substrate. The specific heat should depend linearly
on the density in the coesistence region, varying from the gas value at 
liquefaction to the liquid value at vaporization. While this is not 
exactly the picture of puddles on otherwise empty surfaces proposed 
in Ref.~\onlinecite{sato},
undeniable similarities exist, most notably the density range where the
effect takes place. Consideration of inhomogeneities and/or defects of 
the substrate in the theoretical model may bring numerical and experimental 
evidences in closer agreement.
\\ \indent
It is worth mentioning that the Variational Monte Carlo (VMC) calculation of Ref. \onlinecite{bjl}, making use of the
He-graphite potential of Ref.~\onlinecite{bjl0}, reported evidence for a self-bound 
liquid.  Without questioning the quality of the work of Ref.~\onlinecite{bjl}, or of the richness of the phase diagram of $^3$He 
adsorbed on graphite posited therein, certainly worthy of further investigation, the 
existence of a self-bound liquid can be dismissed on tecnhical grounds. 
For, the energy at zero density is an exact result, while at finite density
the FNDMC energy is by construction lower than the VMC energy (using the same
model Hamiltonian and trial function); since we find that the
FNDMC EOS $E(\rho)-E(0)$ is a non-negative function, so must be the VMC EOS.

\section*{Acknowledgments}
This work was supported in part by the Italian MIUR through PRIN 2011 and by the Natural Science and Engineering Research Council of Canada. 
We acknowledge the CINECA and the Regione Lombardia award $\text{UltraQMC}$, under the LISA initiative, for the availability of high-performance computing resources and support.
One of us (MB) gratefully acknowledges the hospitality of the International School for Advanced Studies in Trieste (Italy).


\begin{thebibliography}{99}
\bibitem{note18}
The choice of a rather small number of $^3$He
atoms is motivated by the simulations of G4He substrates with active $^4$He
in Section IV.b, which require very large numbers of particles at low
$^3$He density.
\bibitem{notafit}
In order to have a smooth correction as a function of density,
we use $\alpha=0.7$ which is nearly optimal for all
densities up to $\rho=0.020$.
\bibitem{notatstep}
The time step error crucially depends on the quality of the trial function.\cite{umrigar93}
A more flexible term than the simple $U_1$ of Eq.~(\ref{eq_psi}) to represent lateral
modulation for corrugated substrates would reduce the strong time step dependence
observed in Fig.~\ref{fig_tau}.
\bibitem{nota} The use of the HFDHE2 presumably explains why
in Ref.~\onlinecite{quasi2d} no gas-liquid coexistence is found for the Mg substrate,
even applying with the less stringent requirement of a negative compressibility
instead of the local minimum criterion followed in Ref.~\onlinecite{quasi2d}.
\end{thebibliography}
\end{document}